# An Auction-based Coordination Strategy for Task-Constrained Multi-Agent Stochastic Planning with Submodular Rewards

Ruifan Liu, Hyo-Sang Shin, Binbin Yan, and Antonios Tsourdos

*Abstract*—In many domains such as transportation and logistics, search and rescue, or cooperative surveillance, tasks are pending to be allocated with the consideration of possible execution uncertainties. Existing task coordination algorithms either ignore the stochastic process or suffer from the computational intensity. Taking advantage of the 'weakly coupled' feature of the problem and the opportunity for coordination in advance, we propose a decentralized auction-based coordination strategy using a newly formulated score function which is generated by forming the problem into task-constrained Markov decision processes (MDPs). The proposed method guarantees convergence and at least 50% optimality in the premise of a submodular reward function. Furthermore, for the implementation on large-scale applications, an approximate variant of the proposed method, namely Deep Auction, is also suggested with the use of neural networks, which is evasive of the troublesome for constructing MDPs. Inspired by the well-known actor-critic architecture, two Transformers are used to map observations to action probabilities and cumulative rewards respectively. Finally, we demonstrate the performance of the two proposed approaches in the context of drone deliveries, where the stochastic planning for the drone league is cast into a stochastic price-collecting Vehicle Routing Problem (VRP) with time windows. Simulation results are compared with state-of-the-art methods in terms of solution quality, planning efficiency and scalability.

## I. INTRODUCTION

Cooperative systems of multiple agents, which features a flexible structure, parallel-processing ability, and scalability, are of great interest, especially for those operating on the unmanned aerial platform [1], such as cooperative surveillance, search and rescue [2][3], border patrolling, etc. Among its various instantiations, there is a specific but widespread category of assigning tasks among team members with a global goal, followed by an independent and possibly stochastic task execution. For example, in deliveries of parcels [4], tasks are allocated to individual vehicles, and the vehicles deliver their allocated items in sequence, without interference from other executors. However, the delivery may subject to stochastic travel delays between destinations. Also, in multi-target tracking [5], agents decide the tracking targets and the best action to track based on the estimation of targe manoeuvring. The cooperative task assignment and decision making must be done considering the stochastic state transitions. To summarize, these domains share some common properties: *i)* agents are functionally independent, except for the task completion constraint; *ii)* decisions are made sequentially while taking uncertainty over their consequences into account; *iii)* there is an opportunity for coordination in advance with prior knowledge of task information.

In [6], this type of problem is viewed as a synergistic combination of two challenges, i.e., task coordination and stochastic planning. Existing approaches are mostly proposed from these two perspectives, either using the deterministic algorithm with greedy selections [7][8] or modelling problems as resource-constrained multi-agent MDPs [10]. The major drawback of deterministic approaches is that they do not account for transition uncertainties, suffering a performance loss in the face of stochastic operation environments. The MDPs provide a rich model for representing uncertainties, however they are not scalable. Despite the fact that they are still struggling to solve the problem of interest, these approaches have their own merits. While approaches derived from the greedy selection take use of the independence amongst task executions to accomplish the conflict-free assignment efficiently and tractably, MDPs enhance the results in consideration of transition uncertainties [11][12]. Motivated by the idea of combining these two methods and built on the previous work in [9], we propose a hybrid algorithm unifying the strengths of these two methods.

To summarize, three key contributions are made by this work to address the above practical task-constrained problems. First, we formulate a generic model for task-constrained multi-agent stochastic planning problems based on MDPs, with the ability to explicitly present task dependencies while taking the stochastic execution process into account. Second, an auction-based

Ruifan Liu is with the School of Aerospace, Transport and Manufacturing, Cranfield University, MK43 0AL, UK (e-mail: ruifan.liu@cranfield.ac.uk).
Hyo-Sang Shin is with the School of Aerospace, Transport and Manufacturing, Cranfield University, MK43 0AL, UK (e-mail: h.shin@cranfield.ac.uk).
Binbin Yan is with the School of Astronautics, Northwestern Polytechnical University, 710072, China (yanbinbin@nwpu.edu.cn)
Antonios Tsourdos is with the School of Aerospace, Transport and Manufacturing, Cranfield University, MK43 0AL, UK (e-mail: a.tsourdos@cranfield.ac.uk).

coordination strategy is proposed to allocate tasks among team members according to a novel score function defined by the task-constrained MDPs. The proposed decentralized coordination strategy is then proved to converge and hold 50% optimality with submodular reward functions. Third, a scalable extension to the auction-based strategy, namely Deep Auction, is proposed by utilizing the neural network approximators, which get trained by the reinforcement learning algorithm and benefits the computing efficiency facilitating implementation on large-scale applications, though at a cost of reducing the optimality.

The rest part of the paper is organized as follows: Section II provides an overview of previous works regarding task allocation and constrained MDPs. Section III introduces the motivations for research and gives a formal definition of the problem. Section IV presents a detailed explanation of the proposed coordination strategy along with its theoretical analysis. This is followed in Section V by an approximate variant of the approach using neural networks. Section VI demonstrates the performance of algorithms through simulation in the context of drone deliveries and conducts the comparison with state-of-the-art approaches. Section VII concludes the contribution of the paper and gives a brief plan for future research.

## II. RELATED WORKS

### A. Constrained Multi-agent MDPs

Constrained stochastic planning, characterized by agent-independent execution and non-overlapping constraints over tasks, can be modelled in a generic form of decentralized MDPs or a specific case of constrained multi-agent MDPs (CMMDPs) [10]. Decentralized Markov decision processes (Dec-MDPs) provide a standard framework for modelling multiagent stochastic planning problems, which specify how a stochastic environment behaves when interacting with a set of decision-makers. Despite the powerful modelling ability, the original formulation of Dec-MDPs considers all dependencies among agents, which makes the problem NEXP-hard and significantly impacts the scalability [13]. To release the computational burden, researchers explored the trait of specific applications, and proposed MDP variants with various hypotheses, such as transition-independent Dec-MDPs [14][15], Group aggregated Dec-MDPs [16], and SI-Dec-MDPs that exploit the sparsity of agent interactions [17].

Task/resource-constrained multiagent MDPs, recognized as weakly-coupled MDPs, can be simplified by decomposing into the individual concurrent process with cooperative actions and merging solutions to form the solution of the original problem [18]. A typical hybrid way to address these resource-constrained MDPs is presented by *Meuleau et al.* [19], who incorporates an online coordination scheme with an offline phase where a value function is computed to define the expected reward from a specific state. Following a similar idea, *Boutilier et al.* [20] propose the budgeted MDPs defining the value function and policy with a separate budget parameter, and then piece together the solution by solving the budget allocation problem. However, two drawbacks exist in these approaches: *i)* the complexity of computing the value function depends on the dimension of states, which can suffer with a large number of budget levels or when it is applied to task-constrained problems; *ii)* the online allocation is addressed in a centralized manner with poor scalability and high communication cost. *Agrawal et al.* [6] alternatively formulate task-constrained MDPs in a task-dependent manner, which facilitates its dual decomposition, and then solve it by a greedy heuristic method GAPS. The combination of the greedy heuristic approach and the MDP formalism can also be found in some other applications, such as decentralized planning for non-dedicated agent teams [12]. Its posterior optimality guarantee is exploited with submodular rewards [15].

### B. Multi-agent Coordination Strategy

Various coordination strategies are implemented to generate a joint policy in the context of the MDP decomposition using different parameters as the criterion, either in a centralized or decentralized manner. As mentioned above, *Meuleau et al.* [19] take a centralized coordination strategy in which actions are sequentially taken in a greedy manner according to the gradient computed by value functions. *Boutilier et al.* [20] instead meet the global resource constraint by solving a *multiple-choice knapsack problem* (MCKP). The concerned task constraint could be satisfied in online coordination by solving a standard task allocation problem. Task allocation is fundamentally a combinatorial optimization problem subjecting to a global objective, recognized as NP-hard. *Heuristic approaches*, e.g., genetic algorithms, ant colony optimization, and market-based methods, and *approximate approaches*, e.g., greedy-based algorithm, are developed allowing for the trade-off between optimality and efficiency. Among them, the approximate approaches enable a certain mathematical guarantee of optimality, if the problem meets certain conditions, e.g., submodularity.

In contrast to the above coordination algorithms with a central node, decentralized strategies are preferred in practical applications, especially for large-scale systems, in terms of the flexibility, communication cost and efficiency of multi-node computing. Auction algorithms, which are related to the sale auctions and contain a set of iterative procedures, are one of the efficient methods for assignment problems. In [21], *Choi et al.* employed the auction approach for task selection and proposed the Consensus-based Bundle Algorithm (CBBA), which is an efficient decentralized task allocation algorithm that adopts a consensus mechanism with the guarantee of at least 50% optimality. As variants of CBBA, a sampling pre-process and a lazy strategy are integrated into the algorithm for better scalability [22][23]. These methods show a great performance when tackling large-scale deterministic task assignment problems, yet they do not apply to the concerned multi-agent stochastic planning. In an effort to retain the outcome with transition uncertainties, *Ponda et al.* [24] present a robust extension to CBBA by using the integral of expected margin utility instead. Nevertheless, computing these robust scores takes a huge cost of time, making the allocation process cumbersome and possibly violating the optimality guarantee [24][25].

### III. MATHEMATICAL FORMULATION

This section introduces a motivating scenario of the task-constrained multi-agent stochastic planning problem and then generalizes it into a generic task-constrained MDP model. Necessary definitions and basic concepts related to the subsequent analyses are also provided in this section.

#### A. Motivations

Stochastic planning problem for multi-agent systems with mutual exclusive task constraint arises in many life or engineering scenarios. One typical application is the drone delivery problem where multiple agents (i.e., drones) $i \in Ag$ are sent to accomplish a cooperative mission constituted by a list of heterogeneous tasks (i.e., parcel deliveries) $j \in \Gamma$, and a task assignment scheme $x_{ij} \in \{0,1\}, \forall (i,j) \in Ag \times \Gamma$ is expected to be found optimizing certain mission objectives.

Two constraints shall be met while planning for the drone delivery problem. First is the non-overlapping task constraint that each task can only be assigned to no more than one team member, and the second is that the cumulative number of tasks for each agent is no more than its capability $c_i$. Assignment and routing solutions could be obtained by casting the problem into the classic vehicle routing problem (VRP) and solving it using well-developed heuristic methods. However, compared to ground-based vehicles, drones are severely affected by weather conditions, especially airflow, which might lead to longer time consumption, resulting in an unexpected delay for subsequent delivery tasks. Accommodating this kind of uncertainty in drone planning tends to increase the quality of routes by improving customer satisfaction and stabilizing deliveries.

To model the stochastic decision-making in the drone delivery scenario, the constrained multi-agent Markov decision processes (CMMDPs) provide a natural framework. As execution uncertainties affect the score function of tasks in the same route sequence without interference among drones, the task imposes constraints on the behaviours of drones that are otherwise completely independent. However, compared to a natural sequential decision-making problem, the task coordination problem needs extra effort to fit into an MDP framework. Decisions could be made to the whole task assignment $X \colon \mathbb{R}_{\{0,1\}}^{|Ag| \times |\Gamma|}$ or only one task-agent pair $x_{ij}, i, j \in Ag, \Gamma$ for each step. With the first configuration, the action space will be massive, the dimensions of which are $2^{|Ag| \times |\Gamma|}$, causing difficulties for MDP solving, while the Markov property might be violated in the second configuration, due to the interference among tasks in one sequence. Therefore, instead of these two formulations, we propose an alternative MDP modelling scheme in the following section via the introduction of intermediate states during the mission execution (i.e., system time, drone's state of charge, and drone's location for the delivery problem).

#### B. Task Constrained MDPs

A task-constrained MDP is defined by the tuple $\langle Ag, \Gamma, C, \langle \mathbf{M}_i \rangle_{i \in Ag} \rangle$.

- $Ag$ is the set of agents.
- $\Gamma$ is the set of tasks.
- $C = \{c_i\}$ indicates the capacities of each agent.
- $\mathbf{M}_i$ is the MDP model of agent $i$ which is confined by a global task assignment constraint. It is defined as the tuple $\langle \tau_i, s_i, a_i, p_i, r_i, x_i \rangle$.
  - $\tau_i$ indicates the decision epoch. The value of $\tau_i$ will add 1 after one decision is made to the agent $i$.
  - $s_i \in S_i$ is the factored local state for agent $i$ (e.g., the system time, drone's state of charge, location, etc.).
  - $a_i \in A_i$ is the action for agent $i$, indicating the next task to execute under the current state $s_i$, which is selected from the available task set for agent $i$: $a_i = j, j \in \Gamma_i$. $\Gamma_i$ is defaulted to be $\Gamma$.
  - $p_i = p(s_i'|s_i, a_i)$ represents the probability that the current state $s_i$ is converted to the next state $s_i'$ by acting $a_i$. (e.g., if the state only considers the system time, the next state is then determined by the probability distribution of task duration $t_i(a_i)$: $p(s_i'|s_i, a_i) \sim Pr(t_i(a_i)|i, a_i)$.)
  - $r_i = r(s_i, a_i, s_i')$ is the reward function which is defined by an immediate gain after executing the selected task $a_i$ under state $s_i$.
  - $\boldsymbol{x}_i = \{x_{ij}, \forall j \in \Gamma\}$ denotes the task accomplishment for agent $i$, where $x_{ij} \colon A_i \times \Gamma \to \mathbb{R}_{\{0,1\}}$ has a similar form as the resource consumption function in [10] but is defined as binary indicating whether task $j$ is accomplished by agent $i$.

The goal for solving this MDP is to obtain a joint policy $\pi^*$ that maximizes the joint expected accumulative reward for the agent group:

$$\pi^* = \underset{\pi}{\operatorname{argmax}} E\left[\sum_{i \in Ag} \sum_{\tau_i=0}^{\tau_f} r_i^{\tau_i} | \pi_i\right] = \underset{\pi}{\operatorname{argmax}} \sum_{i \in Ag} V_i(\pi_i, s_0)$$

s.t.
$$\sum_{i \in Ag} x_{ij} \leq 1 \qquad \forall j \in \Gamma$$
$$\sum_{j \in \Gamma} x_{ij} \leq c_i \qquad \forall i \in Ag$$

where $\pi_i$ is the individual policy for agent $i$, $\tau_f$ is the total task number that is executed by the agent $i$. The accumulated reward under the policy $\pi_i$ could also be represented by the expected value $V_i(\pi_i, s_0)$ on the MDP model $\mathbf{M}_i$. The first inequation defines the inter-agent constraint in task allocation, i.e., duplicate selection of tasks is not allowed. The second inequation represents the intra-agent constraint which limits the upper number of tasks executed by agent $i$ to capability $c_i$.

One interesting observation for this task-constrained MDPs is that if given a collection of non-conflict task sets $\{\Gamma_i, \forall i \in Ag\}$, the global task constraint (i.e., the first inequation) is then satisfied and the individual MDP model $\langle \mathbf{M}_i \rangle_{i \in Ag}$ for each agent will be completely independent of the other. Inspired by this insight, we exploit a decomposition approach by reformulating the constrained optimization problem into

$$\pi^*, N_\Gamma^* = \underset{\pi, N_\Gamma}{\operatorname{argmax}} E\left[\sum_{i \in Ag} \sum_{\tau_i=0}^{\min[c_i, |\Gamma_i|]} r_i^{\tau_i} | \pi_i, \Gamma_i\right] = \underset{\pi, N_\Gamma}{\operatorname{argmax}} \sum_{i \in Ag} V_i(\pi_i, s_0; \Gamma_i)$$

where $\Gamma_i$ is the task set allocated to agent $i$, $N_\Gamma$ is a collection of disjoint task sets that $N_\Gamma = \{\Gamma_i : \forall i, j, |\Gamma_i \cup \Gamma_j| = 0\}$.

The problem is then decomposed into two subproblems with the aim of finding the task allocation $N_\Gamma^*$ and the optimal policy $\pi^*$ sequentially: *1)* For any given policy $\pi$, the assignment result $N_\Gamma$ could be decided by solving the conventional task allocation problem, which we refer to as *the Coordination*; *2)* After we got the result from the assignment, the joint policy is then obtained by solving individual MDP problems $\pi_i^* = \operatorname{argmax}_{\pi_i} V_i(\pi_i, s_0; \Gamma_i)$, and merging into the global one. This process is called *Task Routing*.

It should be noted that after decomposition, the optimization problem has been altered from the original one, which might be at a cost of the optimality compromise. Meanwhile, the decomposition benefits the problem solving, which reduces the computational complexity of the problem from NEXP-hard to P-complete.

*C. Preliminary of Submodularity*

This part gives some necessary preliminary concepts for the development and analysis of the proposed task allocation method.

*Definition 1:* (Submodularity) let $N$ be a finite set. A real-valued set function $f: 2^N \to R$ is submodular if, for all $X, Y \subseteq N$,

$$f(X) + f(Y) \geq f(X \cap Y) + f(X \cup Y)$$

Equivalently, for all $A \subseteq B \subseteq N$ and $u \in N \backslash B$,

$$f(A \cup \{u\}) - f(A) \geq f(B \cup \{u\}) - f(B)$$

*Definition 2:* (Monotonicity) A set function $f: 2^N \to R$ is monotone if, for every $A \subseteq B \subseteq N$,

$$f(B) \geq f(A)$$

*Definition 3:* (Matroid) A matroid is a pair $M = (N, I)$ where $N$ is a finite set and $I \subseteq 2^N$ is a collection of independent sets, satisfying:

- $\emptyset \in I$
- $A \subseteq B, B \in I \Rightarrow A \in I$
- $A, B \in I, |A| < |B| \Rightarrow \exists b \in B \backslash A$ such that $A \cup \{b\} \in I$.

We are interested in a ground set that is partitioned as $N = N_1 \cup N_2 \cup \cdots \cup N_k$. The collection of subsets, $I = \{P \subseteq N : \forall i, |P \cup N_i| \leq 1\}$ forms a matroid called a partition matroid. According to the problem scenario, only one task-agent pair from the task-agent pairs is allowed to be selected. If all task-agent pairs are considered as a ground set (i.e. $N: \Gamma \times A$) and each task-agent pair as an element of the ground set. Thus the task allocation problem modelled in this paper could be handled as a maximization problem subject to a partition matroid constraint [23].

A monotone submodular function has good mathematical characters, with which difficulty can be approximated efficiently with a strong quality guarantee [26]. Specifically, a greedy algorithm that incrementally chooses elements by maximizing marginal utility provides the solution with at least $(1 - 1/e)$ optimality or $1/2$ optimality if there is a matroid constraint [26].

## IV. AUCTION-BASED COORDINATION STRATEGY

This section presents a decentralized auction-based task allocation strategy to address the *Coordination* phase of the above described task-constrained MDPs. We first give a brief introduction to the basics of the *Consensus Bundle based Auction method* (CBBA) and its robust variant. Then, taking use of the same framework, a new auction-based coordination strategy is suggested for the task-constrained MDPs with a reformulated score function.

### A. CBBA and its Robust Extension

The CBBA, originally proposed by [21], is a widely-known decentralized task allocation method, which resolves conflict-free assignments by implementing the auction algorithm on networked agents and augmenting it with a consensus protocol. The algorithm consists of two separated phases: *1)* the bundle construction phase, where a bundle of tasks is greedily chosen by each UAV, and *2)* the task consensus phase, where the conflicts of tasks are resolved through communicating and negotiating with neighbours. The algorithm iterates between these two phases until all tasks are settled in a consistent manner.

Further, the discrepancies between the plan and stochastic real-world operating environments motivated robust extensions to CBBA, where the expected reward is instead maximized [24]. For a formal elaboration, let $\boldsymbol{\theta}$ denotes all uncertainty parameters related to the score calculation, the robust version of the objective function is adapted as

$$\max J_i = E_\theta \left\{ \sum_{j=1}^{|\Gamma|} c_{ij}(\boldsymbol{x}_i, \boldsymbol{\theta}) \right\}$$

where $c_{ij}$ is the marginal score of task $j$, given tasks that are already allocated to agent $i$, shown as $\boldsymbol{x}_i$.

It should be noted that except for the effect on the reward score, uncertainties also influence the optimal task order due to the coupling among tasks. All possible task orders must be taken into account when calculating the score of a task bundle, which should be

$$J_i = \max_{\mathbf{p}_i} J_{\mathbf{p}_i} = \max_{\mathbf{p}_i} E_\theta \left\{ \sum_{j=1}^{|\Gamma|} c_{ij}(\mathbf{p}_i, \boldsymbol{\theta}) x_{ij} \right\}$$
$$= \max_{\mathbf{p}_i} \int_{\boldsymbol{\theta} \in \Theta} \left\{ \sum_{j=1}^{|\Gamma|} c_{ij}(\mathbf{p}_i, \boldsymbol{\theta}) x_{ij} \right\} P(\boldsymbol{\theta}) \, d\boldsymbol{\theta}$$

where $\mathbf{p}_i$ indicates the task sequence, following which the task will be executed one by one.

By means of extracting the uncertainty parameters and optimizing the priori-estimated reward score, this extension indeed enhances the robustness of the CBBA algorithm in the uncertain environment. However, analytically computing this robust score is very difficult due to the integral operation of uncertainty parameters and numerous permutations of the task order. For this reason, some researchers adopt a sampling process to approximate these calculations with the assumption that the order of existing tasks is fixed in order to maintain computational tractability [24]. However, the sampling operation remains cumbersome, and meanwhile, it might also violate the *diminishing marginal gain* (DMG) condition of the algorithm convergence. Therefore, a novel task coordination method is proposed in the following section providing a more efficient way to acquire robust scores in the context of the MDP modelling.

### B. The Proposed Coordination Strategy

The basic idea of the proposed task coordination method follows a new reformulation of score functions on the MDP modelling of the problem. In the original allocation problem, the score function defines an expected reward for executing one specific task set, which is used to calculate the marginal score as the coordination bid for consensus. Suppose we have an intended task allocation for one agent at some point during the coordination process, the new score function is defined by the value function with regards to the assigned task set $\Gamma$ and current state $s_\tau$:

$$V_\pi(s_\tau; \Gamma) = E\left[\sum_{t=\tau}^{T} r(s_t, \pi(s_t), s_{t+1}; \Gamma)\right]$$

The maximum objective function is then converted to

$$\max J_i(\Gamma_i) \Rightarrow V_\pi(s_\tau; \Gamma_i)$$

where $V_\pi(s_\tau; \Gamma_i)$ means the expected accumulative reward of the agent under the state $s_\tau$ with the allocated task set $\Gamma_i$. Using the new utility function formulation, the maximum objective is then acquired by solving the value function of MDPs. Given the agent's state $s_i$, the marginal gain of task $j$ is then calculated by the difference between two value functions:

$$c_{ij} = J^*(\Gamma_i \cup \{j\}) - J^*(\Gamma_i) = V_\pi^*(s_\tau; \Gamma_i) - V_\pi^*(s_\tau; \Gamma_i \cup \{j\})$$

Built on the value-formulated score function, the proposed coordination method follows a similar scheme as the CBBA, iterating between the two phases. The main upgradation is in the bundle construction phase, where the task bundle for each agent is constructed by greedily selecting the task with the largest marginal gain defined by value functions. The pseudocode is presented as Algorithm 1. The value for every state in the MDP model is calculated by a policy iteration approach before allocation starts.

**Algorithm1:** Bundle Construction for agent $i$
1. **INPUT:** BUILD BUNDLE $(z_i, y_i, b_i)$, AGENT STATE $s_i$
2. **while** $h$ not empty
3.     **for** $j \in \Gamma \setminus b_i$
4.         $c_{ij} = V_\pi^*(s_i; b_i \cup \{j\}) - V_\pi^*(s_i; b_i)$
5.         $h_{ij} = \mathbb{I}(c_{ij} > y_{ij})$
6.     **end for**
7.     $j^* \leftarrow \mathrm{argmax}_{j \notin b_i} c_{ij} \cdot h_{ij}$
8.     $b_i \leftarrow b_i \cup \{j^*\}$
9.     $y_{ij^*} \leftarrow c_{ij^*}$
10.     $z_{ij^*} \leftarrow i$
11. **end while**

Another advantage of this MDP-formulated method is that all possible orders of tasks in the bundle will be spontaneously considered when calculating the value function. Most of the existing allocation algorithms operate with an implicit assumption that the order of selected tasks keeps fixed when inserting new tasks into the sequence, for ease of computing. This assumption is regardless of the impact of new tasks on the optimal execution order and thus might violate the underlying conditions for optimality guarantees. However, the reformulated score function herein is directly assessed on the task list and is not affected by the task construction sequences. In other words, it has no relationship with the order of task sequence and can be determined as long as the task list is given for a vehicle. Therefore, this formulation helps to further improve the system objective and truly holds the condition for the following optimality analysis.

*C. Convergence Analysis with Submodular Rewards*

Using a similar consensus topology, the sufficient condition of convergence for the proposed method stays the same as CBBA, i.e., the score function must satisfy the *diminishing marginal gain* (DMG) which has the same denotation as submodularity [21]. The submodularity proof of value function is given by *Theorem1*. Prior to the proof, we first define a preliminary condition, termed *Independent Extensibility*.

*Definition 4:* (Independent Extensibility) Let $N$ be a finite set. Assume a stochastic process on a set is defined by a set of stochastic variables $\boldsymbol{\theta}: 2^N \to \{S_t\}$. The stochastic process are independently extensible if, for every $A \subseteq B \subseteq N$, it satisfies

- $\boldsymbol{\theta}(A) \subseteq \boldsymbol{\theta}(B)$
- $P[\boldsymbol{\theta}(B)] = P[\boldsymbol{\theta}(A)] * P[\complement_{\boldsymbol{\theta}(B)}\boldsymbol{\theta}(A)]$

where $P[\boldsymbol{\theta}]$ is the cumulative distributed function of the variable set $\boldsymbol{\theta}$.

The property of independent extensibility could be found in many applications. Using Vehicle Routing Problems (VRPs) to exemplify, driving between two pairs of locations is regarded as independent stochastic processes. In this case, the stochastic process regarding city A and city B follows the time distribution on its arc $t(c_{ab})$. When the set is extended to have city C, we got two more variables on two extra edges, $t(c_{ac})$ and $t(c_{bc})$, which, however, are independent from $t(c_{ab})$. Thus we have $\boldsymbol{\theta}([A,B]) \subseteq \boldsymbol{\theta}([A,B,C])$ and $P(\boldsymbol{\theta}([A,B,C])) = P[t(c_{ab})] * P[t(c_{ac}), t(c_{bc})]$, the stochastic process concerned by VRPs satisfies independent extensibility.

*Theorem1:* Presuming an independently extensible stochastic process $\{S_t\}$, if its reward function $R$ is submodular, its state-value function $V_\pi(s_\tau)$ is also submodular.

*Preliminary:* First, let us parameterize the state transition of MDPs on the premise of task set $\Gamma$ as $\theta_\Gamma$. The value function is then presented as

$$V_\pi^*(s_\tau; \Gamma) = E_{\theta_\Gamma}[R_{\pi^*}(s_\tau; \Gamma, \theta_\Gamma)]$$

where $R_{\pi^*}(s_\tau, \Gamma; \theta_\Gamma) = \sum_{t=\tau}^{\tau+|\Gamma|} r(s_\tau, \pi^*(s_t), s_{t+1}; \Gamma, \theta_\Gamma)$ is the cumulative reward for executing task set $\Gamma$ under the transition $\theta_\Gamma$. It is noted that given the optimal policy $\pi^*$ and state $s_\tau$, $R_{\pi^*}(s_\tau, \Gamma; \theta_\Gamma)$ is also equivalent to the deterministic reward function with respect to task set $\Gamma$ on the condition of the parameter instance $\theta_\Gamma$. Further, assume that the policy $\pi^*$ and state $s_\tau$ stay unchanged during the task coordination, we can get a simplified notation of the value function:

$$V(\Gamma) = E_{\theta_\Gamma}[R(\Gamma; \theta_\Gamma)] \Leftarrow V_\pi^*(s_\tau; \Gamma) \tag{6}$$

As we assume that the process is independently extensible, according to *Definition 4*, the distribution of parameter $\theta_\Gamma$ is not affected by any additional parameters introduced by extra tasks. Considering a superset $\Gamma'$, $\Gamma \subseteq \Gamma'$, the expectancy of $R(\Gamma)$ on $\theta_\Gamma$ equals to its expectancy on $\theta_{\Gamma'}$, as the additional parameter $\theta_{\Gamma'} \backslash \theta_\Gamma$ is independent of $\theta_\Gamma$, which is formally presented as

$$E_{\theta_\Gamma}[R(\Gamma; \theta_\Gamma)] = E_{\theta_{\Gamma'} \backslash \theta_\Gamma} E_{\theta_\Gamma}[R(\Gamma; \theta_{\Gamma'})] = E_{\theta_{\Gamma'}}[R(\Gamma; \theta_{\Gamma'})], \quad \forall \Gamma' \supseteq \Gamma$$

Via this equation, we further obtain a connection between the marginal value function $V(\Gamma') - V(\Gamma)$ and the marginal reward function $R(\Gamma') - R(\Gamma)$:

$$V(\Gamma') - V(\Gamma) = E_{\theta_{\Gamma'}}[R(\Gamma'; \theta_{\Gamma'})] - E_{\theta_\Gamma}[R(\Gamma; \theta_\Gamma)] = E_{\theta_{\Gamma'}}[R(\Gamma'; \theta_{\Gamma'})] - E_{\theta_{\Gamma'}}[R(\Gamma; \theta_{\Gamma'})]$$
$$= E_{\theta_{\Gamma'}}[R(\Gamma'; \theta_{\Gamma'}) - R(\Gamma; \theta_{\Gamma'})]$$

It implies that, just like the state-value function defines the expectation of cumulative rewards on uncertainties, the marginal value function is also the expectation of the marginal reward function on the joint uncertainty parameters $\theta_{\Gamma'}$.

*Proof:* Now let us consider two available task sets $\Gamma_m$ and $\Gamma_n$, $\Gamma_m \subseteq \Gamma_n$. Given the optimal policy $\pi^*$ and the state $s$, the marginal gain of task $j$ with the prior of the task set $\Gamma_m$, according to the expanded expectation formulation of the value function (*), is presented as follows:

$$V(\Gamma_m \cup j) - V(\Gamma_m) = E_{\theta_{\Gamma_m \cup j}}[R(\Gamma_m \cup j; \theta_{\Gamma_m \cup j}) - R(\Gamma_m; \theta_{\Gamma_m \cup j})]$$
$$= E_{\theta_{\Gamma_n \cup j}}[R(\Gamma_m \cup j; \theta_{\Gamma_n \cup j}) - R(\Gamma_m; \theta_{\Gamma_n \cup j})]$$

Similarly, the marginal gain with respect to the task set $\Gamma_n$ is presented as:

$$V(\Gamma_n \cup j) - V(\Gamma_n) = E_{\theta_{\Gamma_n \cup j}}[R(\Gamma_n \cup j; \theta_{\Gamma_n \cup j}) - R(\Gamma_n; \theta_{\Gamma_n \cup j})]$$

As the reward function $R$ is assumed to be submodular, i.e., $R(\Gamma_n \cup j) - R(\Gamma_n) \leq R(\Gamma_m \cup j) - R(\Gamma_m)$, its expectation on the same parameter distributions is going to be submodular, thus:

$$V(\Gamma_n \cup j) - V(\Gamma_n) \leq V(\Gamma_m \cup j) - V(\Gamma_m)$$

∎

### D. Optimality Analysis

According to [27], for problems that maximize a monotone submodular function subject to a matroid constraint, the greedy algorithm is guaranteed to produce a solution that is bigger than 50% of the optimal solution.

*Theorem1* has indicated the submodularity of the objective function with the assumption of the submodular deterministic reward function. The score function is also shown to be monotone since accumulative rewards are not going to decrease when more tasks are included in the bundle. Further, the non-overlapping task constraint could be considered as a partition matroid where all task-agent pairs are defined as a ground set $N: \Gamma \times Ag$ and the collection of subsets, which is limited to at most one element from the same subset. Overall, the concerned task coordination problem satisfies the conditions listed in [27] for the optimality guarantee with the greedy algorithm. The CBBA holds the identical solution with the greedy algorithm when the reward function is DMG. Accordingly, the proposed task coordination strategy retains the guarantee of 50% optimality. Noted that the optimality guarantee is refer to the optimal expected reward rather than actual execution rewards as uncertainty parameters will not be known until tasks are executed.

In summary, the proposed task coordination algorithm, in the context of MDP modelling, designs a new utility function that aims to optimize the expected accumulative reward. The algorithm is expected to enhance the robustness under uncertain task execution while maintaining the convergence, as well as the prior optimal guarantee of 50%, when applied in a decentralized manner.

## V. APPROXIMATION VIA NEURAL NETWORKS

While the coordination strategy presented in the last section provides an efficient solution, the complexity to solve it still remains NP-hard, which means the computing time exponentially increases along with the problem scale. Recent years have witnessed the great potential of deep reinforcement learning (DRL) to resolve large-scale MDPs [28]. These approximation methods with neural networks, as an enhancement to exact methods, e.g., value iteration (VI) and policy iteration (PI), offer the ability for solving high-dimensional or continuous problems. We thus consider an approximation method via neural networks to generate the planning policy and value function for large-scale task-constrained MDPs.

*A. Deep Auction*

The suggested approximate task coordination method, namely Deep Auction, is inspired by the well-known actor-critic architecture. As demonstrated in Section III, two key points to handle the task-constrained stochastic planning problem are to obtain the optimal individual policy by solving local MDPs and to resolve the task allocation with value function as criteria. In that case, at least two approximators are required for policy and value approximations. Interestingly, there are exactly two networks in the actor-critic network architecture: a policy network maps the action probability, and a critic baseline is used to calculate the advantage value for the benefit to reduce variances, which thus drives the deep auction coordination method.

To elaborate on how the Deep Auction method works, Fig.1 depicts its working mechanism in a sequence of a training phase and an implementation phase. The framework of the approximator training, shown in Fig.1(a), inherits the learning paradigm of REINFORCE algorithm with baselines, where two neural networks are updated using the *Advantage* error defined by the difference between the accumulative reward from the environment and the value evaluated by the critic. After training, the *Actor* and the *Critic*, both modelled by a *Transformer* network, are able to map the policy $\pi^*$ and value function $V^*$ given a task set $\Gamma_i$ and the agent status $s_i$ observed from the environment. In terms of the implementation, seen in Fig.1(b), the trained critic baseline constitutes an auctioneer that negotiates with other agents using the coordination strategy proposed in Section IV. Coordination bids $c_{ij}$ are calculated by the marginal score obtained using the value estimate network. By the end of the auctioning process, a consensus on the task assignment is acknowledged by all agents with a task set $\Gamma_i$ allocated to the current agent $i$ and then passed to a route planner. The route planner greedily chooses the next action $a_i$ from the intended task set according to the action probability $\pi(s_i, a_i | \Gamma_i)$ mapped by the policy network till all tasks have been accomplished or another task coordination is invoked.

In the proposed deep auctioned coordination method, the two decomposed problems for solving task-constrained MDPs i.e., the *Coordination* and the *Task Routing*, as stated in Section III, are addressed by introducing the policy network and the critic baseline in an approximate scalable way.

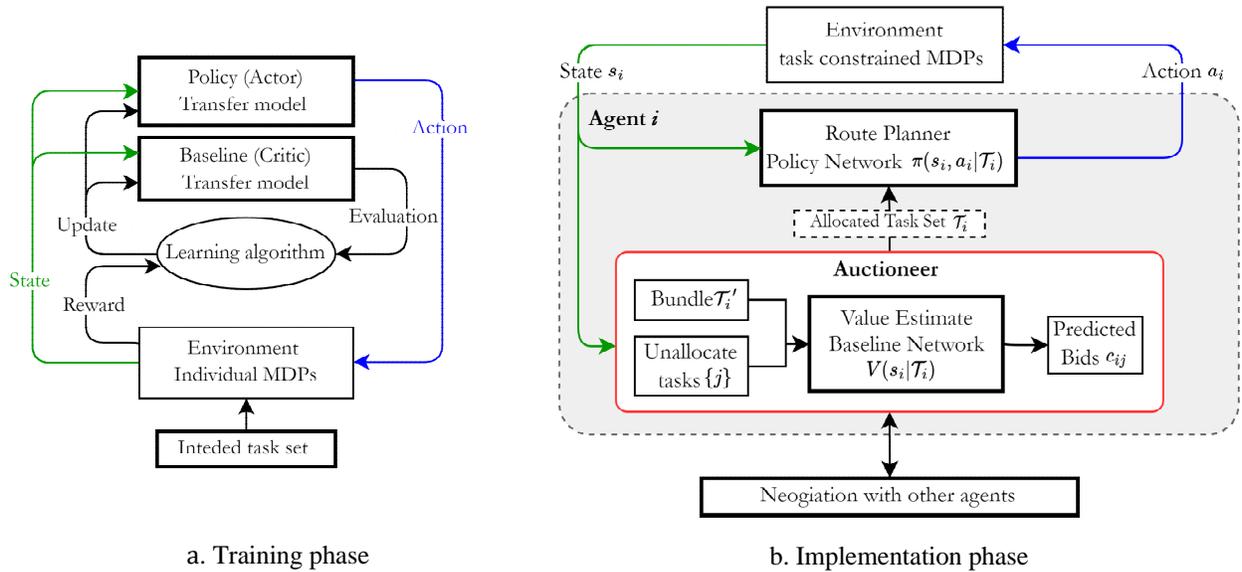

a. Training phase        b. Implementation phase

Fig. 1 The overall framework of the approximate method for solving task-constrained multi-agent stochastic planning problem.

*B. Network Architecture*

The networks employed in the above planning scheme are expected to efficiently extract useful information from task configurations and map the action probability (and expected value function in the critic baseline) from the evolving observations. Thus, the *Transformer* model [30], recognized as a powerful work to address the combinatorial problem with unordered inputs, is adopted here.

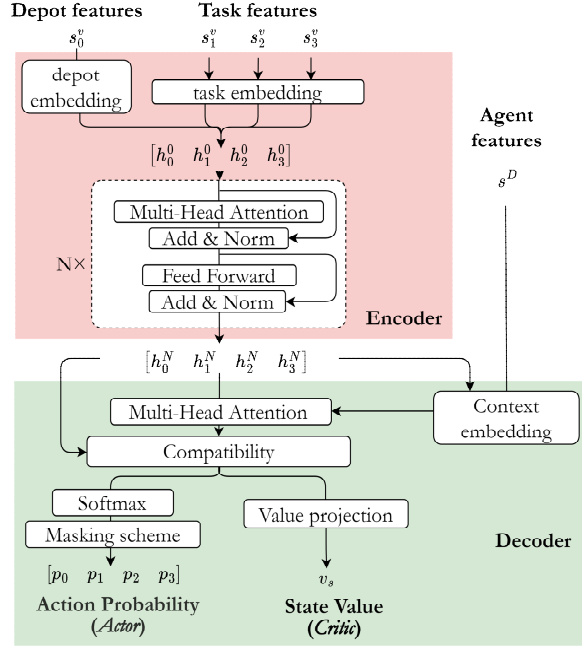

Fig. 2 The neural network architecture. The critic network and actor network share the same architecture except for the output layer for the corresponding output format.

Fig. 2 depicts the architecture of networks, which takes advantage of the encoder-decoder architecture in the Transformer model. The network architecture follows the *Transformer* model presented in [29] and varies slightly in the output layer between the actor network and the critic network. For the actor i.e., the policy network, the output is the probability for all unmasked actions by passing the *compatibility* vector through a SoftMax layer and a Masking scheme. For critic i.e., the baseline, the output is the state value corresponding to the current mission state and agent state, which is decoded by a $N \times 1$ linear layer, where $N$ is the number of tasks.

*C. Training Algorithm*

Networks get trained by the REINFORCE algorithm integrated with the critic baseline. In detail, at the beginning of each episode, a batch of initial states $\{s_0\}_{b \in B}$ and corresponding task sets $\{\Gamma\}_{b \in B}$ are stochastically sampled. Tasks are then sequentially selected from the set according to the probability output of the policy network $\pi(s; \theta)$ and rewards $r(s, a)$ are meanwhile collected. By the end of each episode, we calculated a policy loss $\mathcal{L}$ to train the policy model, which is defined as the *Advantage* error averaged over the batch $B$:

$$\mathcal{L}(B; \boldsymbol{\theta}) = \frac{1}{B} \sum_{b=1}^{N} A(s^b; \boldsymbol{\pi}) = \frac{1}{B} \sum_{b=1}^{N} \left( R(s^b; \boldsymbol{\pi}) - bl(s^b) \right)$$

The *Advantage* error is used for variance reduction and defined by the difference between the accumulative reward and a baseline $bl(s)$. The baseline is calculated by the critic network, which estimates the state-value function, denoted as $\hat{v}(s; w)$ parameterized by $\boldsymbol{w}$:

$$\mathcal{L}(B; \boldsymbol{\theta}) = \frac{1}{B} \sum_{b=1}^{N} \left( R(s^b; \boldsymbol{\pi}) - \hat{v}(s^b; \boldsymbol{w}) \right)$$

The function $\hat{v}$ can be learned from the observation $R(s, \boldsymbol{\pi})$, which is the Monte Carlo estimate of the true value function. Thus, we define a value loss based on the MSE between the estimate function and observations concerning parameters $\boldsymbol{w}$:

$$\mathcal{L}_v(B; \boldsymbol{w}) = \frac{1}{B} \sum_{b=1}^{N} \left( R(s^b) - \hat{v}(s^b; \boldsymbol{w}) \right)^2$$

Standard backpropagation is then adopted to update the policy network and the critic network using the gradient descent method. Beneficial from more efficient training, parameters are shared among these two networks except those of the output projection layer. However, in this case, the learning for the critic network might get distracted by policy errors, as the shared parameters are impacted by the variance of the policy loss due to the greedy sampling method. This does not matter in previous studies, as the critic network is only used to facilitate the training of the actor and will be cut off in the implementation. However, in our study, we are looking for good performance of both two networks. To that end, the training is divided into two phases: *1) Generic Training*: All parameters $(\boldsymbol{\theta}, \boldsymbol{w})$ in both networks get trained. Joint parameters are updated using combined gradients:

$$d\boldsymbol{\theta} = \frac{\partial \mathcal{L}(B; \boldsymbol{\theta})}{\partial \boldsymbol{\theta}} = \frac{1}{B}\sum_{b=1}^{B}\left(R(s^b) - \hat{v}(s^b)\right)\nabla_\theta \log \boldsymbol{\pi}_\theta(s^b)$$

$$d\boldsymbol{w} = \frac{\partial \mathcal{L}_v(B; \boldsymbol{w})}{\partial \boldsymbol{w}} = \frac{1}{B}\sum_{b=1}^{B}\partial \left(R(s^b) - \hat{v}(s^b; \boldsymbol{w})\right)^2/\partial \boldsymbol{w}$$

*2) One-hot Critic Training*: In this phase, the value network is detached from the policy network and updated for a more delicate adjustment on $\boldsymbol{w}$, with parameters in the policy network fixed.

## D. Convergence Guarantee

As mentioned in Section IV.C, the sufficient convergence condition for the decentralized auction-based coordination framework is that the *bids* must satisfy *Diminishing Marginal Gain* (DMG). However, the network-approximated score function might violate the property of submodularity due to the inevitable approximating error using neural networks, even though its MDP form has been proven submodular by *Theorem1*. To resolve this, we implement a *bid wrapping function* proposed by [31] that disguises the actual task scores with DMG *bids* when sharing between agents.

The basic idea behind this wrapping function stems from the key insight that the *bids* agents share must satisfy DMG rather than its actual task scores. Thus, when calculating the marginal scores for all available tasks, $c_{ij} = V(\boldsymbol{b}_i \cup \{j\}) - V(\boldsymbol{b}_i), \forall j \in \Gamma \backslash \boldsymbol{b}_i$, each of them is wrapped with the function:

$$c'_{ij} = \min(c_{ij}, y_{ik}), \qquad \forall k \in \boldsymbol{b}_i$$

where $c_{ij}$ is the actual marginal gain, $c'_{ij}$ is the wrapped coordination gain, and $y_{ik}$ is the existing bid for all tasks in its bundle. The bid wrapping function ensures that the new bids are not greater than the bids already in the bundle, and thus will not affect their allocations. For further details and proofs of convergence, please refer to [31].

## VI. COMPARISON WITH BASE METHODS

The proposed work builds on multiple base research and features with advantages on the trade-off question between computational expenses and robustness against uncertainties. In addition to Section IV, where we demonstrated its improved robustness against uncertainties in the operating environment, in this section we will discuss several other properties of the proposed method, mainly from the perspective of the optimality, and time and space complexity, in comparison with the CBBA and the Multi-agent Reinforcement Learning approach.

The first upper side of the proposed methods is the reduction of computing amount when agents are constructing their task bundles, which benefits from the new-formulated non-ordered score function. In the original CBBA, tasks are sequentially inserted into the path with the best location gained by a trial loop of all possible positions [21]:

$$c_{ij}[b_i] = \max_{n \leq |p_i|} S_i^{p_i \oplus_n \{j\}} - S_i^{p_i}, \quad \forall j \notin b_i$$

where the trial number $|p_i|$ depends on the length of the path, increasing linearly with problem dimensions. In contrast, in the proposed method, bids are calculated using a value function defined on the non-ordered task set, thus exempted from the need of finding the best insertion positions. Formally, given the total number of tasks $n$, the depth of the bundle $k$, and the number of agents $a$, we summarize the total amount of score calculation for using the proposed auction method, CBBA, robust CBBA in Table 1.

Table 1 Complexity comparison

| Method | Individual complexity | Total complexity | Individual complexity (k=n) | Total complexity (k=n) |
|---|---|---|---|---|
| Proposed method | $O(nk)$ | $O(ank)$ | $O(n^2)$ | $O(an^2)$ |
| CBBA | $O(nk^2)$ | $O(ank^2)$ | $O(n^3)$ | $O(an^3)$ |
| Robust CBBA | $O(rnk^2)$ | $O(arnk^2)$ | $O(rn^3)$ | $O(arn^3)$ |

where $r$ is the sampling number of robust CBBA.

An improved optimality guarantee is also facilitated by the proposed non-ordered score function. As we previously noted, there is an unnoticeable assumption in the CBBA process and most of its variations for computational tractability, that the path is always fixed when additional tasks are added to the sequence. However, for some mission profiles, especially those with uncertain executions, the optimal order of actions may alter. Ignoring the influence of additional tasks on optimum execution order reduces solution optimality, resulting in a 50% guarantee relative to a weakened optimal value. Instead of defining the score on a path, the suggested method calculates the marginal gain directly on the task bundle, taking into account all permutations of the task bundle, truly holding 50% optimality for the problem.

Finally, the proposed decomposition of Multiagent MDPs (M-MDPs) greatly reduces the complexity of solving M-MDPs by using a predefined task bundle depth $k$. In the proposed work, M-MDPs is fractioned by individuals and utilized to compute bids for auction-based coordinates. Due to that, the constructed MDPs only attempt to solve the problem of dimensions up to the depth of the bundle $k$ instead of the total problem size $n$. For example, if there are $a = 10$ agents serving $n = 50$ customers, for an relatively even allocation the task number for each agent is restricted to $k = 10$. In this case, the dimension of intended problems is 1~10, instead of 50. This characteristic is very imperative for the implementation of large-scale applications. In the integrated M-MDPs, the dimensionalities of the joint action and state space increase exponentially as the problem size increases, though its worst-case complexity is P-complete [33]. It is reported in [32] that one training episode takes 3 mins for solving VRPs of 10 customers while 22 mins/epoch is required for 50 customers. In contrast, the proposed decomposition strategy will keep the training time constant up to 3 mins/epoch regardless of the overall issue size.

## VII. CASE STUDY ON DRONE DELIVERIES

To test the performance and expected properties of the proposed auction-based coordination method, we implement it in the context of the drone delivery problems, which is cast into stochastic price-collecting VRPs with time windows operated on the unmanned aerial platform. In this section, we start by introducing the definition of this problem and how we generate its mission instances. Then we discuss the performances of using the proposed auction-based coordination method in comparison to state-of-the-art. Demonstrations are conducted in terms of the exact method for small-scale problems and the approximate extension to larger problems.

### A. Drone Delivery Problem with Time Windows

Suppose a drone delivery scenario where there is a list of customer requests $j \in \Gamma$ and a fleet of drones $i \in \mathcal{Ag}$ is sent from the depot $(x^0, y^0)$ to serve the customers and finally returns to the depot after visiting all the tasks. The corresponding price $r_j$ is collected from the customer $j$ if any of the drones arrive at its location $(x^j, y^j)$ during the time window $[t_r^j, t_d^j]$, otherwise a pending cost $c_{pen}$ is imposed. In this logistic problem, optimizing the delivery flow indicates maximizing the rewards collected from customer requests while minimizing the loss of failed tasks.

According to the described delivery scenario, we independently generate sets of instances for different problem dimensions using the mission parameters shown in Table 2 and Table 3 for performance evaluation. Some of the mission settings are adapted from [32].

Table 2 Generic parameter values or distributions used to generate instances of drone delivery.

| Parameter notation | Value or distribution |
|---|---|
| Planning horizon (min) | $H = 480$ |
| Number of customers | $N_{small} = [2,3,4,5]$ <br> $N_{large} = [10,20,50]$ |
| Number of drones | $M_{small} = 2$ <br> $M_{large} = [\frac{N}{5}]$ |
| Expected drone flight speed (km/min) | $V = 1$ |
| Depot location (km) | $x^0, y^0 \sim \mathcal{U}(0,100)$ |
| Customer locations (km) | $x^j, y^j \sim \mathcal{U}(0,100)$ |
| Customer price | $r^j = 1$ |
| Penalty for failed customer | $c_{pen} = 1$ |

Table 3 Parameter distributions related to time windows.

| Notation | Value or distribution |
|---|---|
| Is customer constrained by TM? | $\delta_{TW}^j = \mathcal{B}(p_{TW})$ <br> where $p_{TW} \sim \mathcal{U}(0.25, 0.5, 0.75, 1.0)$ |
| Customer ready time (min) | $t_r^j \sim \mathcal{U}(0, t_{max}^j)$ <br> Where $t_{max}^j = H - \frac{dist(0,j)}{V} - \tau^j$ |
| Customer time window width (min) | $w_{TW}^j \sim \mathcal{U}(30, 90)$ |

| Customer service duration (min) | $\tau^j \sim \mathcal{U}(10,30)$ |
|---|---|
| Customer due time (min) | $t_d^j = t_r^j + w_{\text{TW}}^j$ |

Due to the impact of airflows, the time cost on arcades between customers is shown to be stochastic considering the wind-sensitive property of drones. We simulate the stochastic travel times by sampling the drones' speed from a normal distribution, shown in Table 4.

Table 4 The distribution of drone speed

| Notation | distribution |
|---|---|
| Drone flight speed (km/min) | $V \sim \mathcal{N}(\mu_V, \sigma_V^2)$ Where $\mu_V = 1$, $\sigma_V^2 = [0, 0.05, 0.1, 0.2]$ |

### B. Simulations with Small-scale Problems

In the first simulation, we test the coordination algorithm proposed in Section IV on a set of small-size drone delivery problems (task number is 2~5). The simulation result is compared to the original CBBA and the sampling robust extension of CBBA [24]. A Sample process of $N = 1000$ are used in the sampling robust CBBA to approximate the expected score function. A submodular wrapper is also adopted in the robust CBBA method for convergence.

Suppose that there is a list of 2~5 customer requests and a group of 2 drones is sent from one common depot to these requests for the delivery mission. For the evaluation purpose, we randomly generate 100 mission instances of each problem dimension with parameters following Table 2 and Table 3 and then solve it using the above three coordination methods. The assignment results are validated with the environment that subjects to stochastic travel times following the distribution shown in Table 4. Validation is conducted for 100 rounds. The expected cumulative rewards and validation rewards of all three methods are presented in Table 5. Computing times are shown in Table 6.

Table 5 Total task scores: Expected vs. Actual

| Mission setting \ method | | Proposed auction method | | CBBA | | Robust CBBA | |
|---|---|---|---|---|---|---|---|
| Dro. count | Cust. count | Exp. reward | Act. reward (finish rate) | Exp. reward | Act. reward (finish rate) | Exp. reward | Act. reward (finish rate) |
| 2 | 2 | 1.9 | 1.9 (97.5%) | 1.8 | 1.864 (96.6%) | 1.86 | 1.866 (96.65%) |
| 2 | 3 | 2.8474 | 2.843 (97.38%) | 3.0 | 2.773 (96.2%) | 2.881 | 2.816 (96.93%) |
| 2 | 4 | 3.931 | 4.0 (100.0%) | 4.0 | 3.825 (97.08%) | 4.0 | 4.0 (100%) |
| 2 | 5 | 4.9388 | 4.928 (99.28%) | 5.0 | 4.844 (98.44%) | 4.975 | 4.947 (99.47%) |

Table 6 Running time of three methods for solving 1000 validation instances of each problem size

| Method \ Cust. Count | 2 | 3 | 4 | 5 |
|---|---|---|---|---|
| Proposed Auction | $1''$ | $2''$ | $4''$ | $5''$ |
| CBBA | $4''$ | $8''$ | $17''$ | $25''$ |
| Sampling CBBA | $3'35''$ | $8'57''$ | $18'50''$ | $29'33''$ |

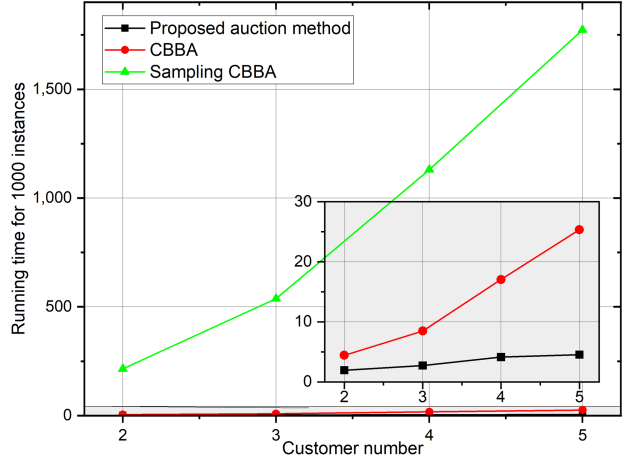

Fig. 3 Running time for all three methods as the customer number changes

Table 5 demonstrates better robustness of the proposed Auction-based method and the robust sampling CBBA against stochastic edge costs. As indicated in the table, the original CBBA has the highest expectation for planning results, even 100% accomplish rates for instances of $N = 3,4,5$, but suffers a performance degradation (to about 97% success rate) when executed in stochastic environments. In contrast, the proposed auction method and sampling CBBA coordinate which use more precise scores though in two distinct ways both result in a greater completion rate (about 1% increase) and a reduced deviation from predicted performance.

While the sampling CBBA performs comparatively to the proposed method, it costs much longer time due to the sampling process, as shown in Table 6. Further, as seen in Fig. 3, the running time for CBBA and robust CBBA present a trend of exponential growth as the customer count increase, while the proposed auction method increases linearly with respect to the problem size, implying a better efficiency on large-scale problems.

*C. Comparative Study via Deep Auction*

To train the two neural networks presented in Deep Auction, we built a simulated environment by which we can evolve the mission state and vehicle state, and get rewards back to update the policy network and the critic network. Instances are generated for training and testing following the parameter scheme in Table 2 and Table 3.

**Training over varied dimensions:** It is noted that for a generalization ability over problem dimensions in value estimate and route planning, datasets for training and testing are constituted by instances of varied dimensions $N \sim \mathcal{U}(1, k)$, where $k$ is the bundle depth defined by the proposed auction-based coordination method.

**Network setup:** For the structure of *Transformer* networks, the depot embedding and task embedding are both one-layer element-wise linear projections with dimension 128. The multi-head attention network consists of 8 heads computing key vectors and value vectors of dimension $d_k = d_v = 16$. There are $N = 3$ multi-attention modules in a row for the encoder network. In the feed-forward layer, the node features are passed through node-wise projections of one hidden sublayer with the ReLu activation and 512 hidden neurons. The Adam Optimizer is used to train networks with a constant learning rate $\alpha_\theta = 10^{-4}$ for the policy network and a constant learning rate $\alpha_w = 10^{-5}$ for the critic network.

We run training for 100 epochs for the generic policy training and additional 50 epochs for one-hot value training. In one epoch, we process 1.28 million instances in 2500 iterations with a batch size of 512. Each epoch takes up to 1 min using the Tesla A100 GPU. Learning curves are presented in Figure 4. All curves integrate data from two learning phases: generic learning for the first 100 epochs and value learning for the 100-150 epochs.

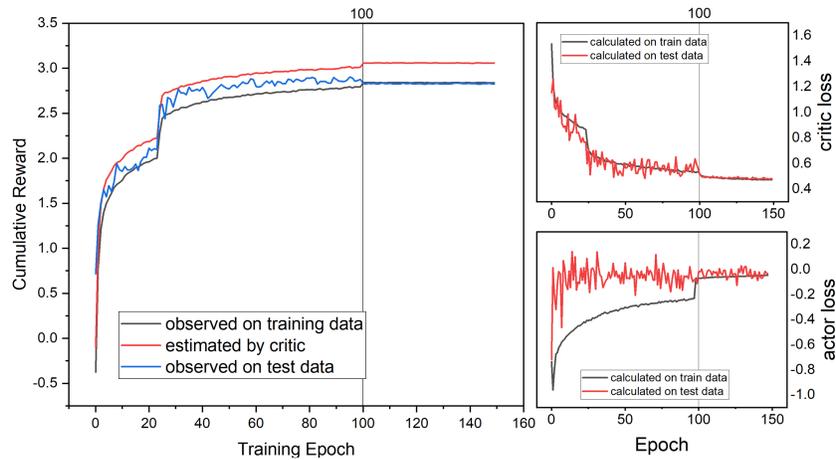

Fig. 4 Learning curves of training actor and critic networks: (left) cumulative reward on training and test datasets per epoch; (centre) policy loss on training and test datasets per epoch; (right) value loss on training and test datasets per epoch.

One interesting points are observed from the figure. First, the cumulative reward obtained from the test dataset gives a better performance than those from the training dataset before the 100th epoch. This phenomenon is reasonable, as the agent stochastically selects its action according to the policy probability at the training phase for the sample diversity while at the testing, actions are chosen greedily for better performance. After 100 epochs, when we fix the policy parameters and use a greedy decoder strategy for both testing and training, the cumulative rewards match rather well even without the overfitting issue. Nevertheless, all curves shown in Fig. 4 demonstrate a convergent training process finalized with stable performance regarding the cumulative reward, actor loss and critic loss.

With the well-trained networks, the proposed deep auction-based approach is evaluated over 3 different mission dimensions $N = [10, 20, 50]$ in comparison with three baselines. The first baseline is the original CBBA, which is decentralized deployed as our method but uses deterministic reward functions. The second baseline considers a hybrid approach that integrates the CBBA task coordination method and the proposed DRL route planner, partially implementing the approximators present in Section V, termed CBBA-DRL. A centralized DRL-based approach, namely MARDAM, proposed in [32], is also included in the comparison. To the best of our knowledge, there is no decentralized instantiation of DRL on routing problems yet due to its computational complexity, so we refer to this centralized approach as a representative from the DRL domain.

Simulation results for solving deterministic instances are shown in Table 7, where we assume that delivery processes are identical with the ones planned, where the drone keeps a constant speed, and the time cost on transport is exactly as expected. For these cases, among the decentralized approaches, including Deep Auction, CBBA, and CBBA-DRL, the CBBA provides best planning solutions for each problem dimension. When using the proposed Deep Auction method, there is a performance decrease of about 2%. Among all methods under consideration, the centralized MARL, i.e., MARDAM, provides the highest quality of solution, which benefits from its global scope of planning.

Table 8 compares the computing time of the proposed Deep Auction method and the CBBA for solving 1000 validation instances of each dimension. Indicated from the table, CBBA consumes longer time to solve the problems than the proposed Deep Auction, and it is noted that the differences become larger as the dimension increases, as shown in Figure 5. The proposed Deep Auction shows a more moderate growth than the CBBA does, which is consistent with the complexity analysis in Section VI and further exemplify the better scalability of the proposed approach.

Table 7 Comparison of cumulative rewards for solving deterministic drone delivery problems with time windows of different dimensions. The value is averaged over 1000 randomly generated instances.

| Mission setting \ method | | Deep Auction | | CBBA | | CBBA-DRL | MARDAM |
|---|---|---|---|---|---|---|---|
| Dro. count | Cust. count | Exp. reward | Act. reward | Exp. reward | Act. reward | Act. reward | Act. reward |
| 2 | 10 | 8.5477 | 8.42 | 8.55 | 8.55 | 8.28 | 9.020 |
| 4 | 20 | 18.396 | 18.04 | 18.78 | 18.78 | 17.21 | 19.080 |
| 10 | 50 | 48.108 | 46.93 | 48.24 | 48.24 | 41.29 | 48.070 |

Table 8 Comparison of the total coordinating time required by the proposed method and the original CBBA for solving 1000 validation instance of drone delivery problems of different dimensions.

| Method \ dim. | Cust. Count = 10 | Cust. Count = 20 | Cust. Count = 50 |
|---|---|---|---|
| Deep Auction | 2′23″ | 12′49″ | 2h2′32″ |
| CBBA | 5′24″ | 54′43″ | 16h28′32″ |

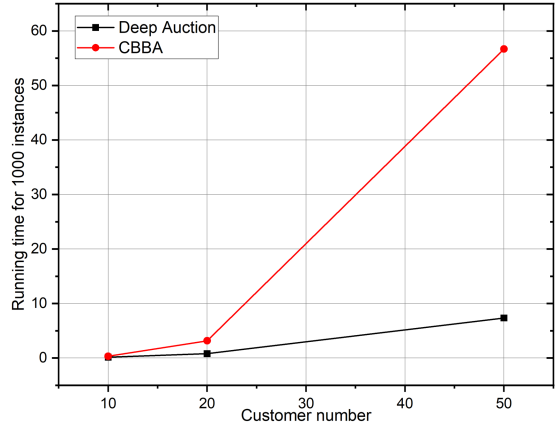

Fig. 5 Running time for Deep Auction and CBBA as the customer number changes

Table 9 Comparison of cumulative rewards for solving stochastic drone delivery problems with time windows of different dimensions. The stochastic feature of deliveries is emulated by sampling the drone speed from a normal distribution defined by different variances. Larger variances indicate higher randomness. The value is averaged over 1000 randomly generated instances.

| Mission setting \ method | | | Deep Auction | | CBBA | | CBBA-DRL | MARDAM |
|---|---|---|---|---|---|---|---|---|
| Dro. count | Cust. count | $\sigma_V^2$ | Exp. reward | Act. reward | Exp. reward | Act. reward | Act. reward | Act. reward |
| 2 | 10 | 0.05 | 8.291 | 8.499 | 8.55 | 8.216 (-3.33%) | 7.987 | 9.043 (+ 6.40%) |
|   |   | 0.10 |   | 8.448 |   | 7.795 (-7.73%) | 7.689 | 8.992 (+6.44%) |
|   |   | 0.20 |   | 8.165 |   | 6.758 (-17.23%) | 7.434 | 8.678 (+6.28%) |
| 4 | 20 | 0.05 | 17.437 | 17.923 | 18.78 | 17.7 (-1.24%) | 16.448 | 18.864 (+5.25%) |
|   |   | 0.10 |   | 17.805 |   | 16.603 (-6.75%) | 16.109 | 18.761 (+5.37%) |
|   |   | 0.20 |   | 17.181 |   | 13.936 (-18.89%) | 15.027 | 18.366 (+6.90%) |
| 10 | 50 | 0.05 | 46.072 | 46.886 | 48.24 | 44.799 (-4.45%) | 39.629 | 47.746 (+1.83%) |
|   |   | 0.10 |   | 46.676 |   | 41.558(-10.96%) | 38.765 | 47.582 (+1.30%) |
|   |   | 0.20 |   | 45.466 |   | 34.275(-24.61%) | 35.967 | 46.909 (+3.17%) |

Second series of simulations are conducted regarding the performance in a stochastic environment. Variances of flight speed $\sigma_V^2 = 0.05, 0.1, 0.2$ are added to the environment, indicating different levels of uncertainties. Simulation results are depicted in Table 9.

If we compare rewards in the table 9 vertically, the DRL-series methods (incl. Deep Auction and MARDAM) perform stably against different speed variances, in contrast with the significant performance degradation using deterministic CBBAs as the variance increases. We attribute the increased robustness to the new formulated score function in the Deep Auction and the observation evolvement in MARDAM.

Second, comparing the rewards horizontally, the proposed Deep Auction gains a better performance compared to the CBBA, while being inferior to the centralized MARDAM method. The inferior performance is within expectation, as centralized methods have the global situation awareness and are more likely to choose better actions than decentralized approaches. However, as previously noted, centralized methods have inherent limitations, due to the central-node dependency and the high communication requirement, as well as the poor scalability which is also demonstrated by the simulation results: the MARDAM losses its superiority incrementally as the problem size increases.

At last, in terms of CBBA-DRL, which is an additional reference incorporating the CBBA for task coordination and a DRL planner for task routing, its performance is poorer than the original CBBA with small variances of 0.05 and 0.1, and surpasses the CBBA with $\sigma_V^2 = 0.2$. This demonstrates the robustness of the DRL planner under uncertainties, though there is a significant negative impact of inconsistent coordination method and route planner.

## VIII. Conclusion

This paper presents two decentralized auction-based approaches addressing the task coordination problem within multi-agent systems while considering the possible operation uncertainties. Through casting the problem as task-constrained MDPs, the task dependency due to an exclusive constraint is despatched from M-MDPs and then resolved by adopting the auction-based coordination method. For these multi-agent stochastic planning problems, the suggested technique resolves the trade-off concern between computational tractability and solution quality. Further, this algorithm is proven to convergent and can reach a posterior optimality guarantee of at least 50% with submodular reward functions. The second algorithm, namely Deep Auction, is an approximate modification of the first suggested auction-based coordination system that includes two neural-network approximators to facilitate large-scale implementations. By theoretical analysis, these two proposed algorithms achieve more robustness and less computing complexity, compared to the original CBBA with respect to the score function calculation and to the centralized MARL in terms of the problem dimensions. Finally, a case study of drone delivery with time windows is studied. The simulation results demonstrate the theoretical benefits of the recommended methodologies.

For the future investigation, the current research will be extended from two aspects. First, though numerical simulations herein are delicately designed for training and validation, more empirical data still benefits the technology consolidation, making it even closer to the real-world scenarios. We would like to gather more experimental data or at least construct finer environments using micro-simulation or discrete-event simulation to train the networks and validate them on more realistic scenarios. Second, while focusing on the robustness against stochastic operations, this paper did not explore the performance of proposed technologies on handling dynamic task lists. Rapid-changing lists of pending tasks might pose challenges demanding a quick replanning ability of the coordination method. Recent work on multi-target tracking [5] predicts neighbours' bids before the coordination. This could be borrowed to reduce the iterations for consensus, facilitating a quicker replanning speed.